\documentclass[12pt]{iopart}

\expandafter\let\csname equation*\endcsname\relax

\expandafter\let\csname endequation*\endcsname\relax

\usepackage{amsmath,amssymb}
\usepackage{graphicx}
\usepackage{bm,ulem}
\usepackage{epstopdf}
\usepackage{color}

\newcommand{\Tc} 	{$T_\text{c}$}
\newcommand{\Tic} 	{$T_\text{iCDW}$}
\newcommand{\Tcc} 	{$T_\text{cCDW}$}
\newcommand{\pdtase}     {2$H$-Pd$_x$TaSe$_2$}

\newcommand{\tase}     {2$H$-TaSe$_2$}
\newcommand{\se}	{$^{77}$Se}
\newcommand{\slr} 	{$T_1^{-1}$}
\newcommand{\slrt} 	{$(T_1T)^{-1}$}

\newcommand{\kk} 	{$\mathcal{K}$}

\begin{document}

\title[Interplay of CDWs, disorder, and superconductivity in 2H-TaSe$_2$]{Interplay of charge density waves, disorder, and superconductivity in 2$H$-TaSe$_2$ elucidated by NMR}

\author{Seung-Ho Baek$^{1,2,\dagger}$, Yeahan Sur$^{3,4}$, Kee Hoon Kim$^{3,4}$, Matthias Vojta$^5$, and Bernd B\"uchner$^{6,7}$}

\address{$^1$ Department of Physics, Changwon National University, Changwon 51139, Korea}
\address{$^2$ Department of Materials Convergence and System Engineering, Changwon National University, Changwon 51139, Korea}
\address{$^3$ Center for Novel State of Complex Materials Research, Department of Physics and Astronomy, Seoul National University, Seoul 151-747, Korea}
\address{$^4$ Institute of Applied Physics, Department of Physics and Astronomy, Seoul National University, Seoul 151-747, Korea}
\address{$^5$ Institut f\"ur Theoretische Physik and W\"urzburg-Dresden Cluster of Excellence ct.qmat, Technische Universit\"at Dresden, 01062 Dresden, Germany}
\address{$^6$ Institut f\"ur Festk\"orper- und Materialphysik and W\"urzburg-Dresden Cluster of Excellence ct.qmat, Technische Universit\"at Dresden, 01062 Dresden, Germany}
\address{$^7$ IFW Dresden, Helmholtzstr. 20, 01069 Dresden, Germany}

\ead{sbaek.fu@gmail.com}

\date{\today}


\begin{abstract}

Single crystals of pristine and 6\% Pd-intercalated \tase\ have been studied by means of
\se\ nuclear magnetic resonance (NMR). The temperature dependence of the \se\ spectrum, with an unexpected line narrowing upon Pd intercalation, unravels the presence of correlated local lattice distortions far above the transition temperature of the charge density wave (CDW) order, thereby supporting a strong-coupling CDW mechanism in \tase. While, the Knight shift data suggest that 
the incommensurate CDW transition involves a partial Fermi surface gap opening. 
As for spin dynamics, the \se\ spin-lattice relaxation rate \slr\ as a function of temperature shows that a pseudogap behavior dominates the low-energy spin excitations even within the CDW phase, and gets stronger along with superconductivity in the Pd-6\% sample. We discuss that CDW fluctuations may be responsible for the pseudogap as well as superconductivity, although the two phenomena are unlikely to be directly linked each other. 
%
\end{abstract}

\submitto{\NJP}
\maketitle



\section{Introduction}

Charge density wave (CDW) order in two dimensions, together with its relationship to superconductivity, has been a central issue in the layered transition metal dichalcogenides (TMDs) \cite{wilson75,castroneto01,rossnagel11}, and even more so owing to remarkable similarities with high-\Tc\ copper-oxide superconductors (cuprates) \cite{chen16}. Electronic phase diagrams in many metallic TMDs suggest that the emergence or enhancement of superconductivity is closely related to CDW order, although the relationship between the two phenomena remains unclear.
Another particularly interesting feature in TMDs is the presence of a pseudogap regime \cite{vescoli98,klemm00,ruzicka01,borisenko08} which involves strange-metal behavior in the normal state. The origin of the pseudogap is often ascribed to a CDW instability \cite{klemm00,bovet04}, an argument also put forward for cuprates \cite{eremin97,li06,wise08}.
Hence the deep understanding of the nature and origin of the CDWs and the pseudogap in TMDs may provide vital clues to understanding the mechanism for high-temperature superconductivity.

\tase\ is one of the intensely investigated TMDs, as it develops a series of fascinating phases: an unusual metallic state with a pseudogap at high temperatures is followed by an incommensurate CDW (iCDW) transition at $T_\text{iCDW}\sim 120$\,K, a lock-in transition into the commensurate CDW (cCDW) at $T_\text{cCDW} \sim 90$\,K, and a superconducting (SC) transition at $T_\text{c}=0.14$\,K \cite{moncton77}. 
The iCDW state may be stabilized by CDW defects termed \textit{discommensurations} which separate commensurate regions \cite{mcmillan75,mcmillan76,suits80,suits81}. In this picture, the lock-in transition could be accounted for by the disappearance of discommensurations. However,
no consensus has been reached on whether the CDW order is of weak-coupling nature, driven by Fermi-surface nesting \cite{straub99,li18,chikina20}, or of more local strong-coupling character \cite{johannes08a,dai14,petkov20}.

Since the two distinct CDW transitions successively occur at much higher temperatures than \Tc, it may be possible to establish a relation between CDW and superconductivity by tuning control parameters such as pressure or doping.
Recently, it has been demonstrated that Pd intercalation leads to the dramatic enhancement of \Tc\ up to $3.3$\,K near an optimal Pd content of $\sim8$\% at which the cCDW completely vanishes while the iCDW transition remains robust \cite{bhoi16}, as shown in Fig.~1(a). %
Being motivated by the strong effect of Pd intercalation on superconductivity and the commensurate CDW in \pdtase, we carried out \se\ nuclear magnetic resonance (NMR) in pristine and 6\% Pd-intercalated \tase. Our NMR data suggest that the major driving force for the CDW formation is strong electron-phonon coupling (EPC). They further suggest that Pd intercalation introduces both changes to the electronic structure and random pinning centers, together being responsible for the strong smearing and suppression of the cCDW transition as well as for the strengthening of the pseudogap behavior on top of the substantial increase of \Tc.


\section{Experimental details}
Single crystals of \pdtase\ [$x=0$ (pristine) and 0.06 (Pd-6\%)] were grown by the chemical vapor transport method as described in detail in Ref. \cite{bhoi16}. It has been confirmed that the Pd intercalation does not alter the $2H$ structure of the pristine compound, as drawn in the inset of Fig.\,1(a).
%
In-plane resistivity $\rho_\text{ab}$ was measured by the conventional four probe technique using a conductive silver epoxy in PPMS\textsuperscript{TM} (Quantum Design). Uniform magnetic susceptibility $\chi$ was measured at 5 T applied along the $ab$ plane in MPMS\textsuperscript{TM} (Quantum Design). Because of the small size of the crystals, the magnetization signal was barely detected in one single crystal so that we stacked 15 single crystals in parallel (a total mass of $\sim 2$ mg) to obtain $\chi$. The temperature dependences of $\rho_\text{ab}$ and $\chi$ for $x=0$ and $x=0.06$ are presented in Figs.\,1 (b) and (c).

\begin{figure}
\centering
\includegraphics[width=\linewidth]{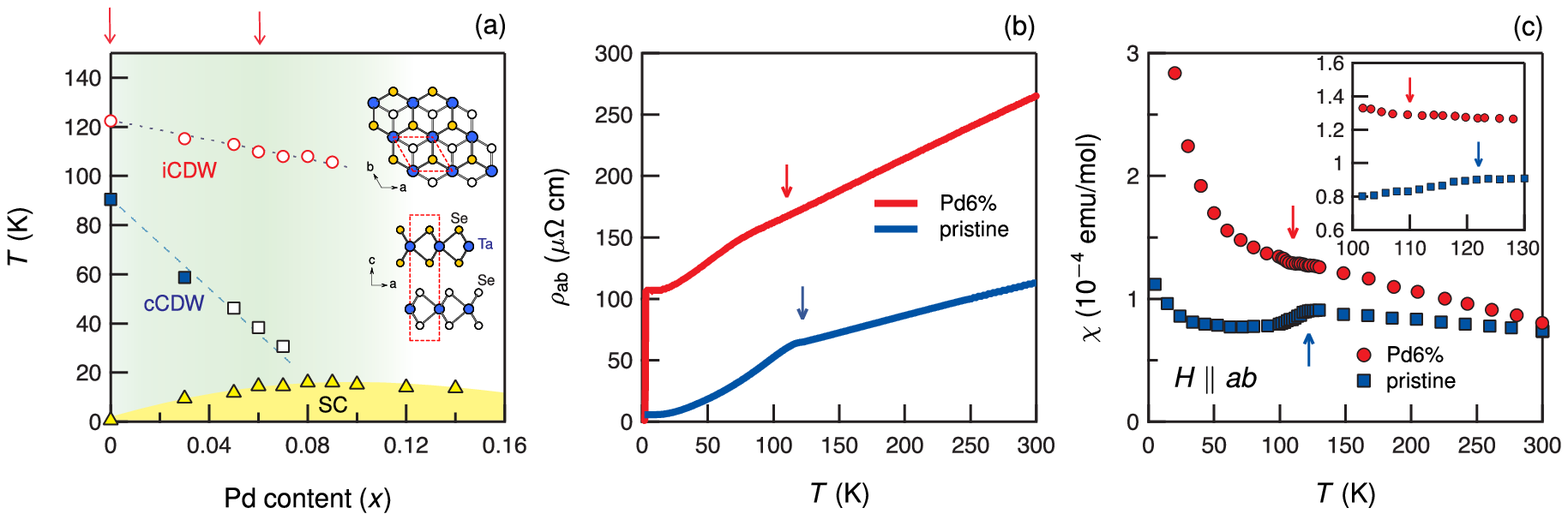}
\caption{
(a) Phase diagram of \pdtase. The CDW transitions \Tic\ and \Tcc, and the superconducting transition \Tc\ (multiplied by 5) are identical with those in Ref.~\cite{bhoi16}. Note, however, that the data points for the cCDW transition for $x>0.04$ are now marked by the empty square unlike in Ref. \cite{bhoi16}, to stress that the transitions smear out, not being well-defined. The red arrows on the top denote the compositions of the samples measured in this work. Inset shows the crystal structure of \tase, where the dashed lines represent the unit cell. (b) Temperature dependence of the in-plane resistivity for pristine and 6\% Pd-intercalated \tase. Note the sharp drop to zero at $\sim 3$\,K for the Pd-6\% sample due to the superconducting transition. (c) Temperature dependence of the uniform magnetic susceptibility measured at 5 T applied along the $ab$ plane. The two arrows in (b) and (c) denote \Tic. The inset enlarges the data near the iCDW transitions. }
\label{phase}
\end{figure}

\se\ (nuclear spin $I = 1/2$) NMR was carried out in pristine and 6\% Pd-intercalated \tase\ single crystals ($\sim 0.7\times 0.5 \times 0.1 \text{ mm}^3$) at an external magnetic field of 15 T, and in the range of temperature 4.2--300 K. The samples were oriented using a goniometer for the accurate alignment along the external field. The \se\ NMR spectra were acquired by a standard spin-echo technique with a typical $\pi/2$ pulse length 2--3 $\mu$s. The nuclear spin-lattice relaxation rate \slr\ was obtained by fitting the recovery of the nuclear magnetization $M(t)$ after a saturating pulse to the following function: $1-M(t)/M(\infty)=A\exp(-t/T_1)$ where $A$ is a fitting parameter that is ideally unity.
In this study, due to the very weak \se\ NMR signal intensity in tiny single crystals, we have focused on the data for $H\parallel ab$ for which the \se\ linewidth is much smaller than that for $H\parallel c$, as the anisotropy of the physical properties is not our main concern of this work.


\begin{figure}
\centering
\includegraphics[width=\linewidth]{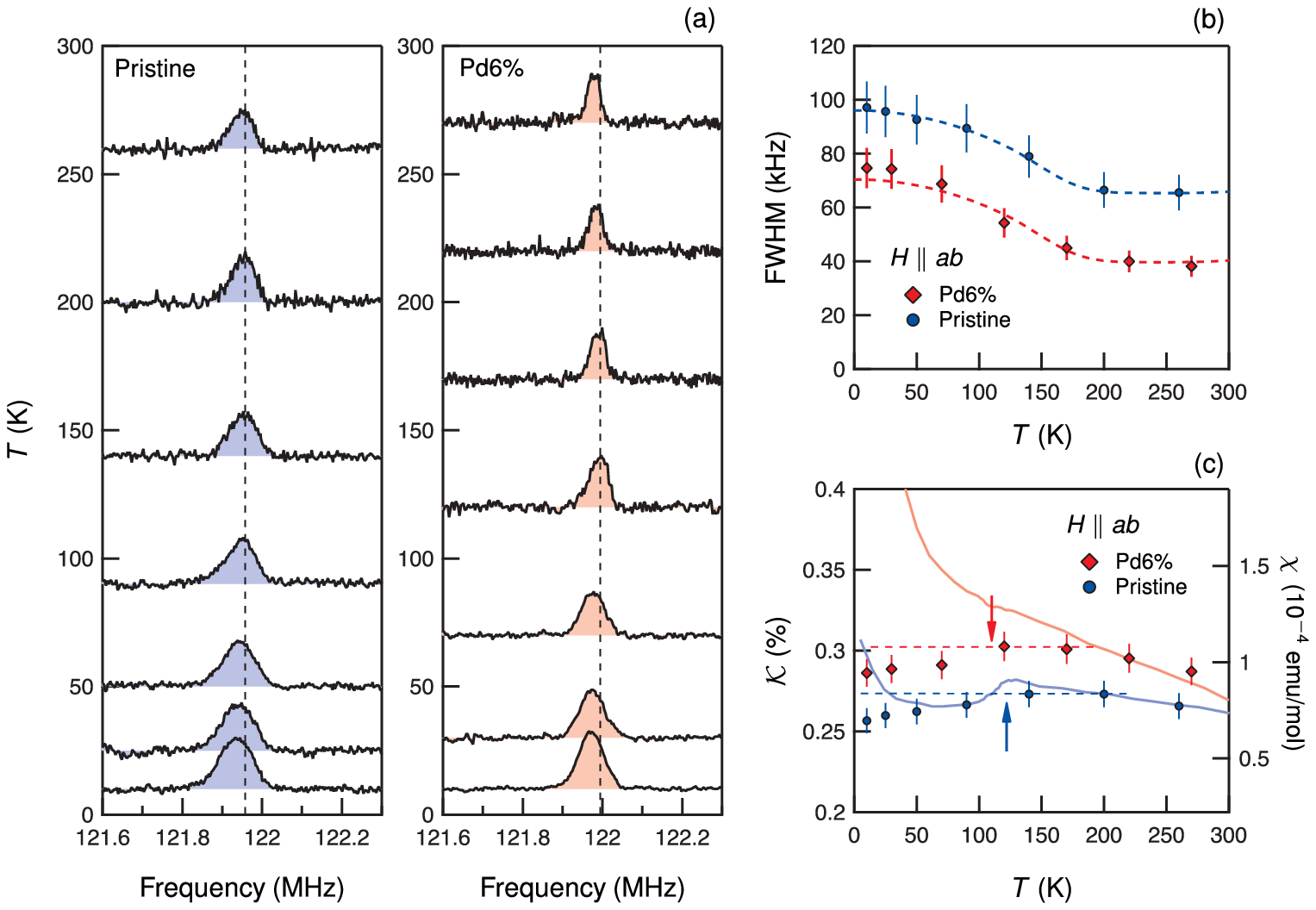}
\caption{(a) Temperature dependence of the \se\ spectrum in \pdtase\ for $x=0$ and $x=0.06$ at $15$\,T applied along the $ab$ plane. The vertical dashed lines were drawn to emphasize the decrease of the resonance frequency at low temperatures. 
(b) Temperature dependence of the \se\ linewidth (FWHM), with the dotted lines being guides to the eye.
(c) Knight shift \kk\ of \se\ as a function of temperature. With 6\% Pd intercalation, \kk\ is notably enhanced preserving its temperature dependence. For the two samples, \kk\ slightly drops below \Tic\ (arrows). The bulk susceptibilities [solid curves, see Fig.~1(c)] are compared. While $\chi(T)$ is consistent with $\mathcal{K}(T)$ for the pristine sample, it is clear that a strong Curie-like tail dominates $\chi(T)$ for the Pd-6\% sample.
}
\label{spec}
\end{figure}

\section{Results}
\subsection{NMR linewdith}

Figure 2(a) shows \se\ NMR spectra in the pristine and Pd-6\% samples as a function of temperature measured at 15 T applied parallel to the $ab$ plane. As shown in Fig.\,2(b), we find that the full width at half maximum (FWHM) of the spectrum, which is a measure of spatial inhomogeneity, 
shows nearly identical temperature dependence for both samples.   
A remarkable and unexpected finding is that the FWHM for the Pd-6\% sample is considerably reduced compared to the pristine one. 
Such a narrowing of an NMR line with doping is extremely unusual, because any dopants inevitably introduce chemical disorder typically leading to a much larger NMR line broadening. In fact, the in-plane resistivity $\rho_\text{ab}$ significantly increases in the Pd-6\% sample,  Fig.\,1(b), reflecting the increased lattice disorder induced by Pd intercalation. Also, the uniform magnetic susceptibility $\chi$, Fig.\,1(c), reveals a much stronger Curie-like tail for the Pd-6\% sample than that for the pristine one, evidencing the increase of localized paramagnetic impurities by Pd intercalation. 

We therefore arrive at the conclusion that the spatial inhomogeneity probed by NMR in pristine \tase\ is largely intrinsic, at least far beyond that originating from impurities. Because of the strong CDW instability in these materials, we think that the intrinsic inhomogeneity arises from local lattice distortions related to CDW phenomena whose amplitude is distributed in space. 
If so, the fact that the FWHM is reduced in the \textit{whole} temperature range investigated, Fig.\,2(b), suggests that local lattice distortions exist even at temperatures far above \Tic, which in turn suggests that the lattice distortions at high temperatures may evolve into the long-ranged CDW state when cooling to below \Tic.
This is clearly inconsistent with a conventional, i.e., weak-coupling CDW mechanism, in which lattice distortions develop only below the CDW transition temperature.
That is, our data indicate that the CDW transition in \tase\ is primarily driven by local EPC \cite{johannes08a,dai14,petkov20} leading to a strong-coupling CDW mechanism with a large fluctuation regime \cite{rossnagel11,mcmillan77,zhu17}, rather than by electronic Fermi-surface nesting of Peierls type \cite{straub99,li18}. 

The question may arise here as to how local lattice distortions become static (at least on the NMR time scale) at $T\gg T_\text{iCDW}$, as they should rapidly fluctuate in time, and how their amplitude is suppressed by external disorder. 
One could argue that the presence of any lattice defects will lead to pinning, such that distortions become inhomogeneous but static, realizing local patches of CDW order. 
Due to the low density of pinning centers, the intrinsic periodicity of the CDW  in pristine \tase\ could be well preserved resulting in the sharp CDW transitions. However, the density of random-field pinning centers rapidly increase upon intercalating Pd, and may strongly compete with the CDW periodicity. This could result in an overall suppression of the distortion amplitude as well as strongly inhomogeneous distortion patterns, which naturally accounts for the nearly indefinable cCDW transition near 6\% Pd content \cite{bhoi16}.
%

\subsection{Knight shift}
The Knight shift \kk, which is equivalent to the local spin susceptibility, of the \se\ spectra is presented in Fig.\,2(c) as a function of temperature. The weak temperature dependence of \kk\ reflects the metallic nature of \pdtase, as \kk\ is  independent of temperature in a Pauli metal.
Nevertheless, \kk\ is slightly enhanced with lowering temperature roughly down to \Tic, indicating that there is a finite non-Pauli contribution to the total spin susceptibility. 
For direct comparison, the bulk susceptibilities $\chi(T)$ are drawn as solid curves in Fig.\,2(c). For the pristine sample, except a Curie-like upturn at low temperatures, $\chi(T)$ is quite consistent with $\mathcal{K}(T)$.  The linear relationship between $\chi$ and \kk\ above \Tic\ permits us to estimate the hyperfine coupling constant, $A_\text{hf}\sim 40$ kOe/$\mu_B$, and the $T$-independent shift $\mathcal{K}_0\sim0.21$ \%, using the relation $\mathcal{K}(T)= A_\text{hf}\chi(T) + \mathcal{K}_0$. 
For the Pd-6\% sample,  $\chi(T)$ turns out to be governed by a strong Curie-like divergence even up to room temperature overwhelming the Pauli contribution, in contrast to $\mathcal{K}(T)$ which shows the almost same temperature dependence as the pristine one. This proves that the Curie-like tails indeed arise from localized impurity moments.

Interestingly, \kk\ looks to be suppressed below \Tic\ for both samples, although there are seldom sufficient data. In great support of this, we note that $\chi(T)$ drops at \Tic\ for the pristine sample. Also, for the Pd-6\% sample, a kink was observed at \Tic\ [see Fig.\,1(c)], which may be identified as a maximum of $\chi(T)$ if the strong Curie background is relatively weak as for the pristine one. The decrease of the spin susceptibility below \Tic\  suggests that the iCDW transition involves a partial gap opening at the Fermi level. Specifically, the Fermi surface nesting may partially contribute to the CDW transition \cite{inosov08}, although it is unlikely the main driving force \cite{johannes08a} as discussed above.
%


%

\subsection{Spin dynamics}
%
We now turn to low-energy spin dynamics probed via the spin-lattice relaxation rate \slr\ which is very sensitive to quasiparticle excitations near the Fermi level in metals.
Figures 3(a) and (b) present \slr\ and \slrt, respectively, as a function of temperature in \pdtase\ ($x=0$ and 0.06) measured at $15$\,T parallel and perpendicular to $c$.
In an ordinary metal, one expects that the relation $T_1^{-1}\propto T$, or $(T_1T)^{-1} \propto n^2(\epsilon_F) = \text{const.}$, holds \cite{bennet}. However, a linear $T$ dependence of \slr\ was not observed, which is better shown in \slrt\ vs.~$T$ plot in Fig.\,3(b).  The rapid decrease of \slrt\ with decreasing $T$ implies a progressive reduction of low-energy degrees of freedom, or the existence of a pseudogap.
In the Pd-6\% sample, \slrt\ decreases faster upon cooling than in the pristine one, i.e., the pseudogap behavior becomes stronger. We also measured the data at low temperatures for $H\parallel c$ to check the anisotropy of the pseudogap. The results, Fig.\,3(b), show that, while \slrt\ is weakly anisotropic, the pseudogap behavior is nearly isotropic being similarly strengthened by Pd intercalation for both field directions. 

Notably, the pseudogap behavior governs the low-energy spin dynamics both above and below \Tic, with no significant change across \Tic. This suggests that the Fermi surface gapping caused by the iCDW transition observed in the Knight shift measurement, Fig.\,2(c), is small compared to the pseudogap. Furthermore, the strengthening of the pseudogap behavior with Pd intercalation 
contrasts with the weakening of lattice distortions and the CDW.
These observations suggest that the pseudogap can be directly linked neither to Fermi surface nesting \cite{borisenko08,chikina20}, nor to static lattice distortions above the transition as expected in a strong-coupling CDW scenario \cite{mcmillan77}. Possibly, the pseudogap is instead related to dynamically fluctuating CDW; such fluctuations of the CDW may also be responsible for the strange metallic behavior characterized by the linear variation of resistivity with temperature \cite{ruzicka01,bhoi16}.

\begin{figure}
\centering
\includegraphics[width=0.8\linewidth]{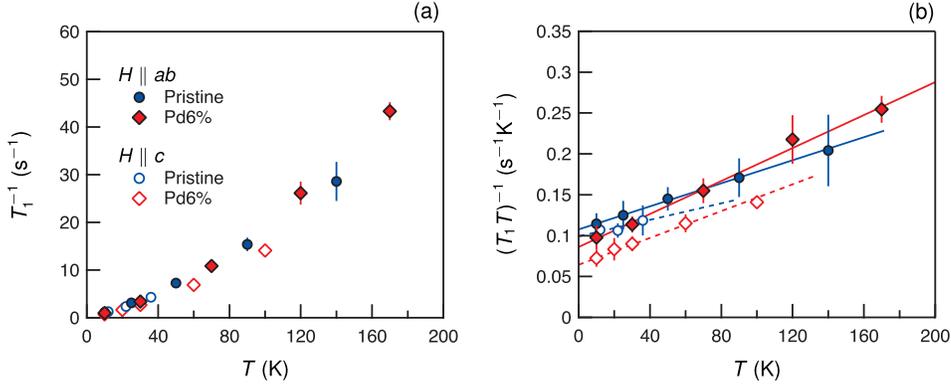}
\caption{Temperature dependence of (a) \slr and (b) \slrt\  at $15$\,T parallel (open symbols) and perpendicular (closed symbols) to the $c$ axis. \slrt\ decreases with decreasing $T$, indicating the presence of the pseudogap. Lines are guides to eye.
}
\label{t1t}
\end{figure}

\section{Discussion}
We now try to synthesize our NMR findings and the phase diagram of Fig.\,1(a) into a coherent picture. Our data suggest that the intrinsically strong EPC generates correlated local lattice distortions in the normal state, ruling much of the phenomena in \tase. 
Pd dopants reduce the distortion amplitude and introduce strong random-field-type pinning which destroys the cCDW state. (Note that we reinterpret the data in Ref. \cite{bhoi16} so that the cCDW transitions above $x=0.04$ are not simply suppressed, but also ill-defined due to the increased density of pinning centers which may hinder the lock-in transition to the cCDW state.) By contrast, the superconducting \Tc\ is dramatically enhanced by increased Pd dopants, in the light of the weakly doping-dependent iCDW transition.

Based on the very different role played by disorder for the CDW transitions and superconductivity, we propose that it may be the \textit{commensurability} of the CDW that competes with superconductivity, rather than CDW ordering itself. Conversely, superconductivity and incommensurate CDW may not be mutually exclusive in nature \cite{wen20}.
Indeed, other superconducting TMDs are also well understood in line with this reasoning. The isostructural 2$H$-NbSe$_2$ and 2$H$-NbS$_2$ \cite{wilson75} in which
the cCDW transition is absent, exhibit $T_c=7.2$\,K and $6.3$\,K, respectively, much higher than $0.14$\,K and $0.8$\,K of the Ta counterparts in which the cCDW is present\footnote{Although the lock-in transition has not been reported for 2$H$-TaS$_2$, the local charge density at $4.2$\,K is found to be commensurate by NMR \cite{nishihara83}. In this respect, the  seemingly competing relationship between superconductivity and the CDW in this material, as discussed in Refs.~\cite{liu21,xu21}, may require a further consideration of the commensurability of the CDW.} 
Furthermore, suppressing \Tic\ in 2$H$-NbSe$_2$ \cite{leroux15} has a negligible effect on $T_c$, in agreement with a weak correlation between iCDW and superconductivity as in Fig.\,1(a).
In contrast, suppressing \Tcc\ either by doping or by external pressure in 1$T$ polytypes mostly induces a substantial increase of $T_c$ --- 1$T$-TaSe$_2$ \cite{liu16a}, 1$T$-TaS$_2$ \cite{sipos08,ang13}, 1$T$-TiSe$_2$ \cite{morosan06,kusmartseva09}, and 1$T$-VSe$_2$ \cite{sahoo20}, as it precisely did in \tase. 


Lastly, the fact that both the pseudogap behavior and superconductivity are boosted by Pd intercalation may suggest that the two phenomena are linked in that CDW fluctuations contribute to both. In this case, the very different temperature scales of pseudogap and superconductivity in \tase\ suggest that the pseudogap in TMDs is a phenomenon distinct from superconductivity, rather than a state of preformed pairs which is often discussed for cuprates \cite{emery95, anderson06, kohsaka08}.

\section{Summary}
We carried out \se\ NMR measurements in pristine and 6\% Pd-intercalated \tase\ single crystals. Our NMR results, combined with resistivity and uniform magnetic susceptibility measurements, strongly suggest that correlated local lattice distortions exist in the normal state as precursor of CDW formation, and thus a strong-coupling CDW mechanism arising from electron-phonon coupling is likely at play in \tase. We argue that the high density of random pinning disorder and modifications to the band-structure caused by Pd intercalation reduce the lattice distortion amplitude and destroy the intrinsic periodicity of the CDW, which accounts for the strong smearing and suppression of the incommensurate-commensurate CDW transition.  
In contrast, 
the pseudogap behavior is much enhanced by Pd intercalation, as is the superconducting transition \Tc. Based on the subtle effects of external disorder, we propose that quenching the commensurability of the CDW may be a crucial factor for the enhancement of \Tc.

\section*{Acknowledgments}

This work was supported by the National Research Foundation of Korea (NRF) grant funded by the Korea government (MSIT) (NRF-2020R1A2C1003817 and NRF-2019R1A2C2090648) and by the Ministry of Education (2021R1A6C101B418). 
MV and BB acknowledge support from the Deutsche Forschungsgemeinschaft through SFB 1143 (project-id 247310070) and the W\"urzburg-Dresden Cluster of Excellence \textit{ct.qmat} (EXC 2147, project-id 390858490).

\section*{References}

\bibliography{mybib}


\end{document}